\documentclass{aastex}

\RequirePackage{color}

\begin{document}
\textwidth=18.5cm
\leftmargin=-1.5cm
\title{The planetary spin and rotation period: A modern approach}
\shorttitle{The planetary spin and rotation period: A modern approach}

\author{A. I. Arbab$^1$, Saadia E. Salih$^2$, Sultan H. Hassan$^1$, Ahmed El
Agali$^1$ and Husam Abubaker$^1$}
\affil{$^1$Department of Physics,
Faculty of Science, University of Khartoum, P.O. Box 321, Khartoum
11115, Sudan}
\and
\affil{$^2$Department of Physics, College of Applied and Industrial
Science, University of Bahri,  Khartoum, Sudan}
\email{aiarbab@uofk.edu}

\begin{abstract}
Using a new approach, we have obtained a formula for
calculating the rotation period and radius of planets. In the
ordinary gravitomagnetism the gravitational spin ($S$) orbit ($L$)
coupling, $\vec{L}\cdot\vec{S}\propto L^2$, while our model predicts
 that $\vec{L}\cdot\vec{S}\propto \frac{m}{M}\,L^2$, where $M$ and $m$ are the central and orbiting masses, respectively.
 Hence, planets during their evolution exchange $L$ and $S$ until they reach a final stability at which $MS\propto mL$, or $S\propto \frac{m^2}{v}$, where  $v$ is the orbital velocity of the planet.
 Rotational properties of our planetary system and exoplanets are in agreement with our predictions.
 The radius ($R$) and rotational period ($D$) of tidally locked planet at a distance $a$ from its star, are related
  by, $D^2\propto \sqrt{\frac{M}{m^3}}\,\,R^3$ and that $R\propto \sqrt{\frac{m}{M}}\,\, a$.
\end{abstract}
\maketitle

\section{Introduction}
Kepler's laws best describe the dynamics of our planetary system as regards to the orbital motion. However, Newton's law of gravitation provided the theoretical framework of these laws. In central potential the orbital angular momentum is conserved. In polar coordinates, the gravitational force consists of the ordinary attraction gravitational force and a repulsive centripetal force. The Newton's law of gravitation has been successful in many respect.
However, this law fails to account for very minute gravitational effect like deflection of light by an intervening star, precession of the perihelion of the planetary orbit and the gravitational red-shift of light passing a differential gravitational potential. Einstein's general theory of gravitation generalizes Newton's theory of gravitational to give a full account for all these observed gravitational phenomena. Einstein  treats these phenomena as arising from the curvature of space. Hence, Einstein's theory has become now the only accepted theory of gravitation.
 The inclusion of energy and momentum of matter (mass) in question leads to the curvature of space, while the inclusion of spin leads to torsion  in space. Einstein's theory deals with matter of the former case, while Einstein-Cartan deals with the latter case. Thus, Einstein space if  torsion free. In classical electrodynamics the spin of a particle is a quantum effect with no classical analogue. However, the spin of a gravitating  object (e.g., planets) is defined as a rotation of an object relative to its center of mass. This is expressed as $S=I\omega$, where $I$ and $\omega$  are the moment of inertia and angular velocity of the rotating object, respectively. The spin is  generally a conserved quantity in physics. Besides the spin,  an object ($m$) revolving at a distant $r$ around a central mass ($M$) with speed $v$ is described by its orbital angular momentum. This is defined  as $L=\vec{r}\times m\vec{v}$. This quantity is also conserved, except when an external torque is acting on the object. In quantum mechanics, the spin  and angular momentum of a fundamental particle are quantized. No such quantization is deemed to exist  in gravitation. To incorporate quantum mechanics  in gravitation we invoke a Planck-like constant  characterizing every gravitational system [1, 2]. This would facilitate a bridging to  quantum gravity that has not yet been uniquely formulated so far.

The spin and orbital angular momentum may couple to each other as the case in the Earth-Moon system. Therefore, neither the spin nor the orbital angular momentum are separately conserved. Their sum is always conserved. A similar coupling occurs in atomic system. For instance, because of the spin of the electron such effect is found to be present in hydrogen-like atoms.

Owing to the existing similarities  between gravitation and electromagnetism, some analogies were drawn which led to
gravitomagnetism paradigm. It is believed that an effect occurring in electromagnetism will have its counter analogue in gravitomagnetism.

In this paper we formulate the proper spin-orbit coupling in a gravitational system, and then deduce a formula for the spin of a gravitating object. This is done by equating the spin-orbit coupling energy to the gravitomagnetic energy. The resulting equation relates the spin of a gravitating object to its orbital angular momentum.
While in standard gravitomagnetism, the gravitational spin-orbit coupling, $\vec{L}\cdot\vec{S}\propto L^2$,  in our model of gravitomagnetism one has $\vec{L}\cdot\vec{S}\propto \frac{m}{M}\,L^2$. This relation suggests a balance equation, $mL\sim MS$. For this reason any orbiting object must spin in order to be dynamically stable. So planets during  their course of evolution exchange $L$ and  $S$, but eventually come to a state of stability. The bigger the planet the larger its spin. Hence,  Jupiter spins faster than other planets in the solar system. Equivalently, the spin $S\propto \frac{Gm^2}{v}$, where $G$ is the gravitational  constant, and $v$ is the orbital velocity. This formula is found to be consistent when applied to our planetary system and exoplanetary system. Astronomers have discovered so far more than 800 new giants planets, but couldn't identify all of their radii and spin
 periods. The present formulation helps identify these latter properties. We consider here all possibilities to account for the observationally  derived data pertaining to the exoplanetary system and their consistency.
\section{The gravitational spin-orbit coupling}
The spin - orbit interaction  resulting  from an interaction of the electron spin with the magnetic field arising from electron motion in hydrogen-like atom is given by
\begin{equation}
U_{SO}=\frac{g_s}{4m^2c^2r^2}\frac{dV}{dr}\,\vec{L}\cdot\vec{S}=-\frac{kZe^2}{2m^2c^2r^3}\vec{L}\cdot\vec{S}\,,
\end{equation}
where $V=\frac{kZe^2}{r}$, $Z$ is the atomic number, $k$ is the Coulomb constant, $m$ is the electron mass, $c$ is the speed of light, $r$ is the radial distance of the electron from the nucleus, and $g_s=2$ is the gyromagnetic ratio.

In gravitomagnetism theory, we have shown that [3],
\begin{equation}
U_{SO-gm}=\frac{\pi g'_s GM^2}{4m^2c^2r^3}\vec{L}\cdot\vec{S}\,,
\end{equation}
$g\,'_s$ is the gravitational gyromagnetic ratio, which corresponds to a gravitomagnetic  energy
\begin{equation}
U_{gm}=-\frac{\pi}{3}\,\frac{GML^2}{2mc^2r^3}\,,
\end{equation}
However,  Einstein's theory of gravitation employing Schwartzchild metric shows that because of space curvature a term of
\begin{equation}
U_{GR}=-\,\frac{GML^2}{2mc^2r^3}\,,
\end{equation}
appears in the total energy of the gravitating object.

Thus, eq.(3) and (4) are very close to each other. This minute difference between the two paradigms should be further explored. Notice that the inclusion of energy momentum tensor in Einstein relativity equations leads to the space curvature, whereas the inclusion of spin would lead to the space torsion. Einstein's general relativity
respects the former but not the latter case. While Einstein
attributed the precession of planets to the curvature of space, we ascribe it to the interaction of the spin of planets with the gravitomagnetic field induced by the Sun in the planet frame of reference.

Assuming the spin-orbit coupling as the one responsible for precession of perihelion of planetary orbits, the spin of a planet of mass $m$ orbiting a star of mass $M'$ can be obtained by equating eqs.(2) and
(3), \emph{i.e.}, spin-orbit interaction energy equals to gravitomagnetic energy, which yields
\begin{equation}
S=\frac{\alpha}{\cos\theta}\frac{m}{M}\,L\,,
\end{equation}
where $L$ is the orbital angular momentum of the orbiting planet\footnote{$L$ is perpendicular to the ecliptic.}, $\alpha=\frac{2g'_s}{3}$, $\theta$ is the angle between $L$ and $S$ directions,  and $M=(M'+m)^\dagger$\footnote{$M\sim M'$} is the total mass of the system (see figure 1). However, the orbital plane of most planets is inclined to their ecliptic with an angle ($i$). We have to decompose $L$ along the direction that is perpendicular to the ecliptic (our reference plane), hence $L\rightarrow L\cos i$ (see figure 1). Therefore, eq.(5) becomes
\begin{equation}
S=\alpha\frac{\cos i}{\cos\theta}\,\mu\,\,L\,,\qquad \mu=\frac{m}{M}\,.
\end{equation}
Equation (6) indicates that for a system of particles each having a mass $m_i$, and  an angular momentum  $L_i$, the the center of mass of the angular  momentum is
\begin{equation}
L_C=\frac{\Sigma_i\,m_i\,L_i}{\Sigma_i\,m_i}\,,
\end{equation}
so that the spin could be related to a center of mass of the angular momenta of the system.

Owing to the apparent analogy between electromagnetism and gravitomagnetism, one has
\begin{equation}
G\rightarrow k\,,\qquad M\rightarrow Ze\,,\qquad m\rightarrow e\,,
\end{equation}
so that eq.(1) would become
\begin{equation}
U\,'_{SO}=\frac{g_g G}{4\mu\,c^2r^3}\vec{L}\cdot\vec{S}\,,
\end{equation}
where $g_g$ is gravitational gyromagnetic ratio analogue. It is not known whether $g_g=2$ or not. According to the standard theory of gravitomagnetism, we can  equate  eqs.(4) and (9) to obtain  $S=\frac{L}{\cos\theta}$ for $g_g=2$. This is however not correct for the planetary system. Equation (9) agrees with eq.(2)  only if $g_g=\pi\frac{M}{m}\, g'_s$.

We remark that several authors have considered the gravitational spin-orbit coupling comparing it with the atomic analogue [4, 5].
 None of them have derived it from first principle, or equivalently didn't show how the gravitational spin magnetic moment is related to the spin.   This is only done in our recent publication [3].
Mashhoon  proposed that the analogy between gravity and electromagnetism dictates that charge, $q\rightarrow -2m$.  Hence, he concluded that the  gravitational magnetic moment due to spin is related to spin by $\mu_{s}=-S$ [6]. In our gravitomagnetic theory, this is however related by  the relation $\mu_g=\frac{M}{2m}\, S$.

Applying eq.(8) in eq.(3) dictates that the curvature term lead to a potential (interaction) energy in the atomic system
\begin{equation}
U_{em}=-\frac{\pi}{3}\,\frac{kZe^2L^2}{2m^2c^2r^3}\,.
\end{equation}
Therefore, one can write the total potential energy for an electron in hydrogen-like atoms in an electric space as \begin{equation}
E=-\frac{kZe^2}{r}+\frac{L^2}{2mr^2}-\frac{\pi}{3}\,\frac{kZe^2L^2}{2m^2c^2r^3}\,.
\end{equation}
Comparing eqs.(1) and (10) reveals that $\vec{L}\cdot\vec{S}=\frac{\pi}{3}\,L^2$ for an atomic system. Thus,
with this understanding the space inside an atom is not flat space (Minkowskian), but follows Schwarzschild pattern, where $\frac{2GM}{c^2}\rightarrow\frac{2kZe^2}{mc^2}$ is the electrical Schwarzschild radius. Hence, the metric for a spherically symmetric distribution of nuclear matter can be written as
\begin{equation}
ds^2=c^2(1-\frac{2kZe^2}{mc^2r})\,dt^2-\frac{dr^2}{1-\frac{2kZe^2}{mc^2r}}-r^2(d\theta^2+\sin\theta^2d\varphi^2)\,.
\end{equation}
One can then interpret the Rutherford-$\alpha$ deflection as a consequence of the electrical curvature inside the atom. This is tantamount to deflection of light by the Sun curvature. The electric Schwarzschild radius is equal to  twice classical  electron radius, $R_s=2r_c=\frac{2kZe^2}{mc^2}$. Similarly, one would expect a photon to be electrically redshifted in an electrical potential of  the nucleus by an amount, $z=\frac{kZe^2}{mc^2r}$ in hydrogen-like atoms. Hence, in an analogous manner when light passe near a central charge it will  experience a redshift. This can be written as, $z=\alpha\,Z\left(\frac{\lambda_C}{r}\right)$, where  $\lambda_C=\frac{\hbar}{mc}$ is Compton wavelength of the electron. Now, when $r=\lambda_C$, then $z_m=\alpha\, Z$. This case represents a  maximal (quantum) redshift. Thus, owing to the one-to-one analogy between gravitation  and electromagnetism, one can use Einstein's general relativity to describe electromagnetic phenomena, and Maxwell's equation to  describe gravitational phenomena.
\section{The planetary spin and radius}
The origin of spin of planets has not been known exactly. One can easily determine the orbital angular momentum of a planet. The spin of a planet however requires knowledge of the planet mass, radius,
its rotation period and its mass distribution inside the planet. Since some planets are solid (rocky) and other are gaseous, it is not easily to identify precisely their internal structure. The former ones have generally higher rotation rate than the latter. However, orbital periods of planets depend on their distance from
the Sun  and the Sun mass only. We provide here a formula for spin or rotational period from its orbital motion only. Or equivalently, we relate the  spin to the orbital angular momentum for the first time in history.

Equation (6) can be  used to express the planetary spin as
\begin{equation}
S=\alpha\frac{\cos i}{\cos\theta}\,\,\frac{G\,m^2}{v}\,.
\end{equation}
This is a very interesting and useful formula that can be used to calculate the spin of a planet without resort to its rotation period and its radius. Delauney and Flammarion related the spin period of a planet to a host of planetary physical characteristics concluding that there is a direct relationship between spin period and mean density [7]. Furthermore, Brosche noticed that some planets in a similar size range had spin angular momenta, $S$, that were
proportional  to the squares of their masses, $m$ [7, 8]
\begin{equation}
S\propto m^2\,.
\end{equation}
Equation (14) agree partially with eq.(13). In rotational dynamics the spin of a rigid body (planet) is defined by
\begin{equation}
S=I\omega\,,\qquad \omega=\frac{2\pi}{D}\,,\qquad I=\lambda\,mR^2\,,
\end{equation}
where $\lambda$ is the coefficient of inertia, $R$ the planet's radius, and $D$ is the rotational period of the planet.
 Equation (13) and (15) states that the radius of the planet is
\begin{equation}
R=\left(\frac{\alpha\cos
i}{2\pi\lambda\cos\theta}\,\frac{G\,m\,D}{v}\right)^{1/2}\,.
\end{equation}
Using eqs. (6) and (15) one can write
\begin{equation}
\textcolor[rgb]{0.00,0.00,1.00}{\frac{D}{P}=\frac{1}{C^2}\,\left(\frac{R}{a}\right)^2\frac{1}{\mu}}\,,\qquad
C=\left(\frac{\alpha\,\cos
i\sqrt{1-e^2}}{\lambda\,\cos\,\theta}\,\right)^{1/2}\,,
\end{equation}
where $P$ and $a$ are, respectively, the orbital period and the semi-major axis of the orbiting planet. Equation (17) can be written as
\begin{equation}
\textcolor[rgb]{0.00,0.00,1.00}{D=\frac{1}{C^2}\,\left(\frac{4\pi^2}{GM}\right)^{2/3}\left(\frac{1}{\mu}\right)\,
R^2\,\,P^{-1/3}\,.}
\end{equation}
An educated guess can relate $\alpha$ to the ellipticity (flattening/oblateness) of the planet, or to the eccentricity of the orbit. If the value of $\alpha$ is not universal for all planets, we suggest that it will depend on some geometrical factors related to a given planet. This particular relation will require more analysis that can be tackled in future work. Using eq.(15), the rotation spin rate can be written as
\begin{equation}
\omega=\alpha\frac{\cos
i}{\cos\theta}\left(\frac{\mu\sqrt{G\,M\,a(1-e^2)}}{\lambda\,R^2}\right)\,\,,
\end{equation}
Equation (19) can be written as
\begin{equation}
\omega=\alpha\frac{\cos i}{\cos\theta}\frac{L}{I_\lambda}\,\,,\qquad
I_\lambda=\lambda MR^2\,,\qquad L=\sqrt{G\,M\,m^2\,a(1-e^2)}\,.
\end{equation}
The radius of a planet that is tidally locked to its star, \emph{i.e.}, $P=D$, is given by (\emph{see eq.}(17))
\begin{equation}
\textcolor[rgb]{1.00,0.00,0.00}{R_t=C\,\sqrt{\mu}\,\,a}\,\,.
\end{equation}
Equation (18) can also be written as, for $P=D$,
\begin{equation}
\textcolor[rgb]{1.00,0.00,0.00}{P_t^2=C^{-3}\,\left(\frac{4\pi^2}{GM}\right)\left(\frac{1}{\mu}\right)^{3/2}\,R_t^3}\,.
\end{equation}
It is of prime interest to mention that a hypothetical satellite that had a circular orbit radius equals to the radius of the planet, $R$, its orbital period $P$  is given by [7]
\begin{equation}
P_R^2=\left(\frac{4\pi}{GM}\right)\,R^3\,.
\end{equation}
Flammarion  calculated $P_R$ values for Earth, Jupiter, Saturn, Uranus and Neptune, by extrapolating Kepler’s Harmonic Law, as applied to their known satellites [7]. Only some of these period are
in agreement with observation.

The radius of a black hole  is related to its mass, $m$,  by
\begin{equation}
R=\frac{2Gm}{c^2}\,.
\end{equation}
Therefore, the gravitational force for such a planet (spinning black hole) is given by
\begin{equation}
F_N=\frac{G\,m\,M}{r^2}=C^2(1-e^2)\,\frac{c^4}{4G}\,\frac{D}{P}\,,
\end{equation}
where $r=a(1-e^2)$. This clearly shows that a spinning black hole will experience a huge gravitational force when orbits any central massive object. It is shown by [10] that the spin of black hole ($m_B$) is given by $S=\chi \,G m_B^2/c$, where $\chi$ is some constant. This agrees with eq.(13) where $v$ replaces $c$ for a black hole.

This force is maximum when the planet is tidally-locked to its star, \emph{i.e.}, $P=D$. Hence, one has
\begin{equation}
F^{\,max.}_N=C^2(1-e^2)\frac{c^4}{4\,G}\,.
\end{equation}
This force is of the order of $\frac{c^4}{G}$. It is however shown that the maximal force in nature is defined by
$F_{max.}=\frac{c^4}{4\,G}$ [1, 2, 11]. It also represents the maximum self-gravitating mass. It is thus interesting to see that the gravitational force arising from this case is of the same order of this maximal force. For a black hole planet of radius $R_p$ tidally-locked with a black hole star with radius $R_s$, one has
\begin{equation}
R_pR_s=C^2\,a^2\,.
\end{equation}
This is an interesting relation connecting the two radii of orbiting black that are tidally-locked to their semi major axis. Moreover, it is clear that the existence of such a system awaits the future astronomical exploration.
\section{Results and discussions}
We consider here the planetary system, Jupiter satellites, and Saturn satellites. The constant $C$ is calculated using eq.(18) and Tabulated in Tables 3 and 4. The average value of $C$ are  0.089, 0.077, and 0.068, for  Jupiter satellites, Saturn satellites, and planetary system, respectively.
Notice that for exoplanet and asteroids the constant $C$ takes the average values 7 and 0.5, respectively. The higher values of $C$ for asteroids may be attributed to the uncertainty associated with  the observational data related to them. Table6 can be used to identify exoplanets that are tidally locked  by comparing the values of $C$, for a given system, with that of the Moon.  Since our Moon is tidally locked, and well-known, we can consider $C=0.041$ for tidally locked planets.  With this value, we complete table 7. Assuming the exoplanetary system is similar to our planetary system, we suggest that $C=0.1$, as evident from table 1. With this value we calculate the day for some exoplanets as shown in table 7. Notice that Mercury is very closed to tidally-locked system envisaged in Tables 2 and 6.
\section{Conclusion}
Einstein's general theory of relativity modifies the Newton's law of gravitational by adding an extra term that Einstein attributed to the space curvature. We have shown in this work that this term could also arise from the spin-orbit interaction of spinning gravitating (planets) objects with the  gravitomagnetic field.  This assumption yields spin values for the planetary systems that are in agreement with observations. The equations associated with spin are then used to identify and calculate the astronomical data related to the newly discovered planets (exoplanets).
\section*{References}
\hspace{-0.95cm} $[1]$ Arbab, A. I., \emph{Gen. Relativ. Gravit.}, 36, 2004, 2465.\\
 $[2]$ Arbab, A. I., \emph{African J.  Math. Phys.}, 2, 2005, 1.\\
 $[3]$ Arbab, A. I., \emph{J. Mod. Phys.}, 3, 2012, 1231; Arbab, A. I., \emph{Astrophys. Space Sc.}, 330, 2010, 61\\
 $[4]$ Lee, T. -Y., \emph{Physics Letters A}, 291, 2001, 1.\\
 $[5]$ Faruque, S. B., \emph{Physics Letters A}, 359, 2006, 252.\\
 $[6]$ Mashhoon, B., \emph{Physics Letters A}, 173, 1993, 347.\\
 $[7]$ Hughes, D.W., \emph{Planetary and Space Science}, 51, 2003,  517. \\
 $[8]$ Brosche, P., \emph{Icarus}, 7, 1967, 132.\\
 $[9]$ Data are obtained from http://www.wolframalpha.com,   http://nssdc.gsfc.nasa.gov/planetary/factsheet/, and http://exoplanet.eu\\
  $[10]$ Kramer, M., \emph{General Relativity with Double Pulsars}, SLAC Summer Institute on Particle Physics (SSI04), Aug. 2-13, 2004, pg.1.\\
 $[11]$ Massa, C., \emph{Astrophys. Space Sci.}, 232, 1995, 143.

\newpage

\begin{figure}[!htb]
\centering
\includegraphics[scale=.7]{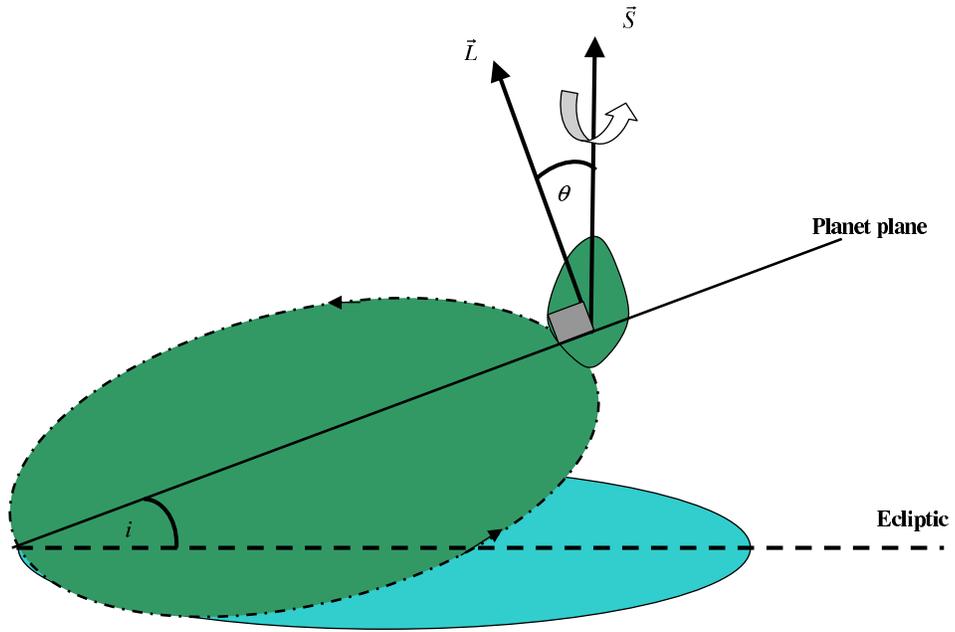}
\caption{The inclination angle ($i$), the angle ($\theta$), the spin ($S$), and angular momentum $L$.}
\label{fig:digraph}
\end{figure}
\begin{table}[p]
\centering
\begin{tabular}{|l|c|c|c|c|c|c|c|c|}
  \hline
 Name & $m (\times 10^{24}\rm kg)$  & $P (day)$ &   $D\, (hr)$ &      $e$  & $a\,(au)$&  $R \,(km)$ & $C$ \\
   \hline
     \hline
Mercury& 0.3302  & 87.969 &       1407.6 &  0.2056 &    57.91 &    2439.7 &  0.036390953 \\
  \hline
 Venus & 4.8685 & 224.701 &      5832.5 &     0.0067 & 108.21 &   6051.8 &  0.009877991 \\
\hline
 Earth & 5.9736 & 365.256 &      23.9345 &    0.0167 & 149.6 &    6378.1 &    0.13530133 \\
  \hline
 Mars & 0.64185 & 686.98 &      24.6229 &    0.0935 &   227.92 &    3396.2 &    0.195060979 \\
  \hline
Jupiter & 1898.60 & 4332.59&     9.925 &     0.04093&  778.57 &    71492 &    0.087422946\\
  \hline
 Saturn & 568.46 & 10759.22 &  10.656 &    0.0489 &    1433.53&    60268 &   0.111249286  \\
  \hline
Uranus  & 86.832 & 30685.40 &   17.24 &    0.0565 &   2872.46&     25559&    0.079987318  \\
\hline
Neptune & 102.43 & 60189  &    16.11 &      0.0457 &  24764&     4495.06 &  0.066062714 \\
\hline
Pluto& 0.01305 & 89866   &      153 &       0.244671&  5874&    5874 &      0.082682797 \\
\hline
  \hline
\end{tabular}
\caption{The planetary system primary data}
\end{table}

\begin{table}[p]
\centering
\begin{tabular}{|l|c|c|c|c|c|c|c|c|}
  \hline
 Name & $m (\times 10^{22}\rm kg)$  & $a\,(\,\rm km)$ &   $P\, (day)$ &      $D(day)$  & $e$&  $R \,(km)$ & $C$ \\
   \hline
     \hline
Moon &  7.3477    & 384400 &    27.321582
 &  27.321582&    0.0549 &   1738.14 &  0.041 \\
  \hline\hline
\end{tabular}
\caption{The Moon (tidally locked to the Earth)  primary data}
\end{table}

\begin{table}[p]
\centering
\begin{tabular}{|l|c|c|c|c|c|c|c|c|}
  \hline
 Name & $m (\times 10^{20}\rm kg)$  & $a\,(10^3\,\rm km)$ &   $P\, (day)$ &      $D(day)$  & $e$&  $R \,(km)$ & $C$ \\
   \hline
     \hline
Io &  893.2   & 421.6  &      1.769138 &  1.769138 &     0.004 &    1821.6 &  0.0116 \\
  \hline
Europa &    480 & 670.9&      3.551181 &    3.551181 &  0.0101 &   1560.8 &  0.0085\\
\hline
 Ganymede &  1481.9 & 1070.4 &   7.1545535 &   7.154553 &   0.0015 &    2631.2 &   0.0051\\
  \hline
 Callisto &    1075.9 & 1882.7 &      16.689018 &    16.689018 &  0.007 &    2410.3 &    0.0031 \\
  \hline
Elara & 0.008    &  11740&    259.6528 &   0.5   &     0.217 &          40 &    0.0695\\
  \hline
 Himalia &  0.095 & 11460&      250.5662 &  0.4  &     0.162    &           85&    0.0483  \\
  \hline
Metis & 0.001 & 128       &      0.294779 &    0.294779 &    0.0002 &        20 &    0.3959  \\
\hline  \hline
\end{tabular}
\caption{Jupiter Satellites  primary data}
\end{table}

\begin{table}[p]
\centering
\begin{tabular}{|l|c|c|c|c|c|c|c|c|}
\hline
Name & $m (\times 10^{20}\rm kg)$  & $a\,(10^3\,\rm km)$ &   $P\, (day)$ &      $D(day)$  & $e$&  $R \,(km)$ & $C$ \\
\hline \hline
Miranda & 0.66&   129.39  &      1.413479 &  1.413479 &     0.0027 &    235.8 &  0.1797 \\
\hline
Ariel &  13.5&   191.02&      2.520379&    2.520379 &  0.0034 &   578.9 &  0.0661\\
\hline
Umbriel & 11.7 &   266.3 &   4.144177&   4.144177 &   0.005 &   584.7 &   0.0514\\
\hline
Titania & 35.2 & 435.91&     8.705872 &    8.705872 & 0.0022 &   788.9 &    0.0244\\
\hline
Oberon& 30.1 &   583.52&      13.463239 &   13.463239   &     0.0008 &          761.4&   0.0191\\
\hline  \hline
\end{tabular}
\caption{Saturn Satellites (tidally locked) primary data}
\end{table}

\begin{table}[p]
\centering
\begin{tabular}{|l|c|c|c|c|c|c|c|c|}
\hline
Name & $m (\times 10^{19}\rm kg)$  & $a\,$(au) &   $P\, (year)$ &      $D(day)$  & $e$&  $R \,(km)$ & $C$ \\
\hline \hline
Ceres & 87&   2.767  &   4.6     &  9.075 &    0.0789 &    487.3 &  3.75\\
\hline
Juno & 2 & 2.669&   4.36&     7.21&      0.2579 &   120  &  6.90\\
\hline
Vesta & 30 &   2.362 &   3.63&   5.342 & 0.0895   &   265 &   4.71\\
\hline
Eugenia & 0.61 &    2.721&   4.49 &   5.699 &  0.0831 &   113 &    13.16\\
\hline
Siwa & 0.15 &   2.734&    4.51 &   18.5   &   0.2157  &         51.5&   6.70\\
\hline
Chiron & 0.4 &   13.633&        50 &   5.9   &   0.3801 &          90&   8.48\\
\hline
Haumea & 41.79 &   43.335&    285.4 &   3.912   &    0.18874 &         718&  6.11\\
\hline
Pallas & 3.18 &  2.7707&      3.62 &   7.8132      &   0.231  &          261&   10.03\\
\hline
Eris & 1.62 &  2.385&     3.68 &   7.14    &   0.231  &    199.8 &   13.18\\
\hline  \hline

\end{tabular}
\caption{Asteroids primary data}
\end{table}

\begin{table}[h]
\centering
\begin{tabular}{|l|c|c|c|c|c|c|}
\hline
Name & $m (M_J)$& $M(M_\odot)$  & $a\,$(au) &    $R (R_J)$ & $P(day)$ &$C$ \\
\hline \hline
Kepler-34(AB) b&    0.22&   2.0687& 1.0896& 0.76&        288.822& 0.0350\\
\hline
Kepler-9 c  &0.171  &1& 0.225&  0.823&      38.9086 &0.0395\\
\hline
KOI-55 c&   0.0021& 0.496&  0.0076& 0.078&  0.34289& 0.0395\\
\hline
HD 97658 b& 0.02&   0.85&   0.0797& 0.262&9.4957    & 0.0329\\
\hline
GJ 3470 b&  0.044&  0.541&  0.0348  &0.376&     3.33714& 0.0453\\
\hline
Kepler-22b& 0.11&   0.97&   0.85&   0.214&  289.9&0.0425\\
\hline
Gl 581 g    & 0.01  & 0.31 &    0.14601&    0.0678& 36.652  &   0.0429\\
\hline
\end{tabular}
\caption{Tidally-locked Exoplanets primary data [9]}
\end{table}

\begin{table}[h]
\centering
\begin{tabular}{|l|c|c|c|c|c|c|c|}
\hline
Name & $m (M_J)$& $M(M_\odot)$  & $a\,$(au) &   $P\, (day)$ &       $R (R_J)$ & $D (day)$ \\
\hline \hline
WASP-10 b & 3.06 & 0.71&   0.0371   &    3.09276     &       1.08&  0.6566 \\
\hline
XO-5 b &    1.077 &0.88  & 0.0487 &     4.18775&    1.03 & 0.4783\\
\hline
WASP-16 b &     0.855 &1.022  & 0.042 &     3.1186&    1.008 &  0.1891\\
\hline
KOI-204 b&  1.02 & 1.19  & 0.0455 &     3.24674&    1.24 &  0.1564 \\
\hline
XO-2 b& 0.62 & 0.98  & 0.0369 &     2.61584&        0.973 & 0.0993 \\
\hline
TrES-1& 0.761 & 0.88  & 0.0393 &    3.03007&        1.099 &  0.1398\\
\hline
WASP-1 b&   0.86 & 1.24  & 0.0382 &     2.51995 &       1.484 &  0.0483\\
\hline
HAT-P-17 b&     0.534 & 0.857  &    0.0882 &    10.3385 &       1.01 & 2.051\\
\hline
WASP-55 b&  0.57 &  1.01 &      0.0533&     4.46563 &       1.3 &   0.1768\\
\hline
WASP-6 b&   0.503 &     0.888 & 0.0421&     3.36101 &       1.224 &     0.0940\\
\hline
55 Cnc e& 0.0263 &      0.905& 0.0156& 0.736546 &       0.194 &         0.00578\\
\hline
OGLE2-TR-L9 b &     4.34 &  1.52 & 0.0308 &         2.48553 &   1.614 &  0.1079\\
\hline
OGLE-TR-10 b &  0.68 &  1.18 & 0.04162 &        3.10129&     1.72 &  0.04368\\
\hline
XO-3 b &    11.79 &1.213 &  0.0454 &    3.19152&    1.217 &  1.802\\
\hline
PSR 1719-14 b &     1 &1.4 & 0.0044 &   0.0907063&   0.4 &  0.000327\\
\hline
WASP-14 b &     7.341 &1.211 &  0.036 &     2.24377&  1.281 &  0.4484\\
\hline
HD 80606 b &    3.94 &0.98 &    0.449 &     111.436&        0.921 &     4444.9\\
\hline  \hline
\end{tabular}
\caption{Some non-tidally locked Exoplanets primary data [9]. The day is
obtained for $C = 0.1$ as can be guessed from table 2 for planetary
system. }
\end{table}

\end{document}